\documentclass[twocolumn, twocolappendix]{aastex631}

\usepackage{amsmath}
\usepackage{xcolor}

\begin{document}

\title{Temporal Variations in Pulsar Spectro-Polarimetry: Findings from millisecond pulsar J2144$–$5237 using Parkes UWL receiver}
\author[0009-0001-9428-6235]{Rahul Sharan}
\affiliation{National Centre for Radio Astrophysics, NCRA-TIFR, Pune 411007, India}

\author[0000-0002-6287-6900]{Bhaswati Bhattacharyya}
\affiliation{National Centre for Radio Astrophysics, NCRA-TIFR, Pune 411007, India}

\author[0000-0002-7122-4963]{Simon Johnston}
\affiliation{CSIRO Space and Astronomy, P.O. Box 76, Epping, NSW 1710, Australia}

\author[0000-0003-2122-4540]{Patrick Weltevrede}
\affiliation{Jodrell Bank Centre for Astrophysics, Department of Physics and Astronomy, The University of Manchester, UK}

\author[0000-0002-2892-8025]{Jayanta Roy}
\affiliation{National Centre for Radio Astrophysics, NCRA-TIFR, Pune 411007, India}

\begin{abstract}
While the temporal variations of the spectro-polarimetric nature of pulsars remains unexplored, this investigation offers significant potential for uncovering key insights into pulsar emission mechanisms, magnetic field geometry, and propagation effects within the magnetosphere. We developed a package for investigating time-varying spectral behavior for full Stokes parameters and demonstrate it on a millisecond pulsar (MSP) J2144$-$5237 in a binary system (orbital period $\sim$ 10 days) using the Parkes UWL receiver. In this study we report rotation measure (RM) variation with orbital phase. We find that the temporal variations in the spectra of Stokes $I$, $Q$, and $V$ are generally correlated throughout the orbit, while Stokes $U$ exhibits intervals of both correlation and anticorrelation with Stokes $I$, depending on the orbital phase. We also provide a Poincar\'e sphere representation of the polarization properties of J2144$-$5237, demonstrating a systematic temporal change of Poincar\'e sphere location for the main component with orbital phase. To our knowledge, this is the first investigation of time-varying properties of the spectro-polarimetric nature of any pulsars or MSPs. Extending this study to probe the spectro-temporal nature of full Stokes data on a larger sample of MSPs or pulsars has the potential to provide vital information on emission mechanisms inside the magnetosphere, interstellar propagation effects, and binary interactions.
\end{abstract}


\section{Introduction}

Pulsars are recognized as coherent emitters with strong polarization, making their polarization profiles valuable for studying their magnetic structure and the interaction between magnetic fields and accelerated plasma. Certain polarization characteristics, such as those described by the rotation vector model, serve as effective tools for testing the presence of a dipolar magnetic field structure \citep{Radhakrishnan_n_Cooke_1969}.\par
Spectro-polarmetric investigation of millisecond pulsars (pulsars with a spin period $<$ 30 ms) are challenging due to their inherently low brightness. Despite this, such studies are crucial for gaining insights into their emission processes, magnetic field structures, and the propagation effects within their magnetospheres. Polarization profiles reveal the structure of MSP emission beams and magnetic field orientations \citep[e.g.][]{Oswald_et_al_2023}. 
Previous studies in the literature like \cite{Spiewak_et_al_2022, Dai_et_al_2015, Yan_et_al_2011, Stairs_et_al_1999} show diverse nature of full Stokes profiles for MSPs.\par
The study of polarization in MSPs mostly relies on folding observational data over time intervals ranging from a few seconds to tens of seconds, depending on the pulsar's signal strength. This time integration is essential for achieving adequate signal-to-noise ratio (SNR) for detailed polarization analysis. 

Temporal and frequency-dependent variations in polarization properties—such as Stokes intensity and polarization angle—help distinguish intrinsic emission characteristics from effects caused by the interstellar medium (ISM) or binary companions. 

We conducted a systematic investigation probing the temporal changes in the MSP spectra which is reported in \citep{Sharan_et_al_2024}. In that study, we report a significant temporal evolution of the in-band spectra for 10 MSPs discovered by the uGMRT (including J2144$-$5237), observed from 2017 to 2023 using band 3 (300$–$500 MHz) and 4 (550$–$750 MHz) of the uGMRT. However, the analysis was restricted to total intensity (Stokes $I$) measurements.\par
Advancements in broadband radio telescopes, such as the Parkes Ultra-Wideband Low (UWL) \citep{UWL_receiver} receiver and the Green Bank Telescope (GBT)\footnote{\url{https://science.nrao.edu/facilities/gbt}}, have significantly enhanced the quality of spectro-polarimetric data. 
These improvements facilitate more detailed studies of spectral and polarization properties of the MSPs, contributing to a deeper comprehension of their emission physics and aiding in precision pulsar timing efforts.\par
To the best of our knowledge, no comprehensive study has yet explored the temporal variation of spectral properties of full set of Stokes parameters of pulsars. Investigating how these spectro-polarimetric parameters evolve over time can reveal changes in the spectral behavior across a wide range of timescales. The timescales over which significant polarization variations occur may offer insights into the underlying emission mechanisms and propagation effects, potentially shedding light on the physical causes behind changes in the observed polarization state.
\par
We conducted a spectro-polarimetric investigation of MSP J2144$-$5237 using Parkes UWL receiver. As part of this investigation, we also examine how the polarization of the radio signal changes across the pulse phase by representing the Stokes parameters ($Q$, $U$, and $V$, following the definitions in \cite{Born_n_Wolf(1980)}) on the Poincaré sphere. This approach provides a straight forward and intuitive way to visualize changes in the polarization state. Similar techniques have recently been applied in studies of fast radio bursts (FRBs) \citep{Dial_et_al_(2025), Bera_et_al_2025}.
The structure of this paper is as follows: in Section \ref{observations}, we present observations and data analysis methods, Section \ref{Results} presents the results, and finally in Section \ref{summary} we summarize the results obtained.

\section{Observations and Analysis}
\label{observations}

The pulsar J2144$-$5237 was observed with the ultra-wide-bandwidth low-frequency (UWL) receiver on the Parkes radio telescope \cite{UWL_receiver}, spanning a frequency range of 704$-$4032 MHz. The observations were recorded with a time resolution of 30 seconds, the frequency channel resolution was 1 MHz, and 1024 phase bins in the folded mode. In this work, we present results from full-Stokes polarimetric observations conducted on 29 May 2023 (for 8.6 hours), on 22 June 2023 (for 2.2 hours), 26 August 2023 (for 9 hours) and 28 May 2025 (for 3.3 hours).\par

\textsc{PSRPYPE}\footnote{\url{https://github.com/vivekvenkris/psrpype}} was used to process initial step of data analysis. \textsc{PSRPYPE} uses \textsc{CLFD}\footnote{\url{https://github.com/v-morello/clfd/releases/tag/v0.4.0}} \citep{ref_CLFD} for RFI removal.
Flux and polarization calibration is carried out using standard procedures as outlined in \cite{Johnston_et_al_2021} and \cite{Sobey_et_al_2021}. A pulsed signal is injected into the feed to measure the relative gains and phases between the two linear probes of the UWL. Flux calibration is performed using observations of the radio galaxy PKS 1943$-$638. Observations of the pulsar J0437$-$4715 are used to determine the polarization calibration model (pcm) as described in \cite{Straten_2013}. These procedures are implemented within the
\textsc{PAC} command of \textsc{PSRCHIVE}\footnote{\url{https://psrchive.sourceforge.net/}} \citep{Hotan_van_Straten_Manchester_2004} for calibration of the data. 
The calibrated data was saved in standard PSRFITS format.

Due to the large data volume, subsequent processing was carried out in segments to accommodate the memory constraints of the computing environment. 
We read the \textsc{PSRFITS} file using python package \textsc{astropy.io.fits}\footnote{\url{https://docs.astropy.org/en/latest/io/fits/index.html}} \citep{astropy:2022} and perform initial data analysis such as dispersion measure (DM) correction, baseline removal , and transforming the data into Stokes parameters. The optimal Rotation measure (RM, in $rad \: m^{-2}$) and the error on it was determined, for time chunks (or subintegrations) of roughly 20 minutes, using a modified version of RM-Tools\footnote{\url{https://github.com/CIRADA-Tools/RM-Tools}} \citep{Ref_RM_tools}. Finally, we have removed the possible contributions of ionosphere, using \textsc{spinifex}\footnote{\url{https://pypi.org/project/spinifex/}} \citep{ref_spinifex}, from the estimated RM values. Internally, \textsc{spinifex} uses publicly available global ionospheric maps in IONEX format\footnote{\url{https://spinifex.readthedocs.io/en/latest/iono_models.html}}, providing TEC (total electron content) for estimating ionospheric RM. The ionospheric RM estimates are subject to accuracy of the available TEC measurements, at the observing site. Then we calculated the spectra for Stokes $I$, $Q$, $U$, $V$, and the linear polarization, $L$=$\sqrt{Q^2 + U^2}$. The 
variation of Stokes  $Q$, $U$ and $V$ with the pulse phase was visualized on the Poincar\'e sphere for each observing epoch. An example of this representation for 22 June 2023, the observing epoch with highest SNR detection, is shown in Figure \ref{All_comp_spectra_fig}. Further analysis is present in Section \ref{RM_orbital_phase}.

\begin{figure}[htbp]
  \centering
  \includegraphics[scale=0.17]{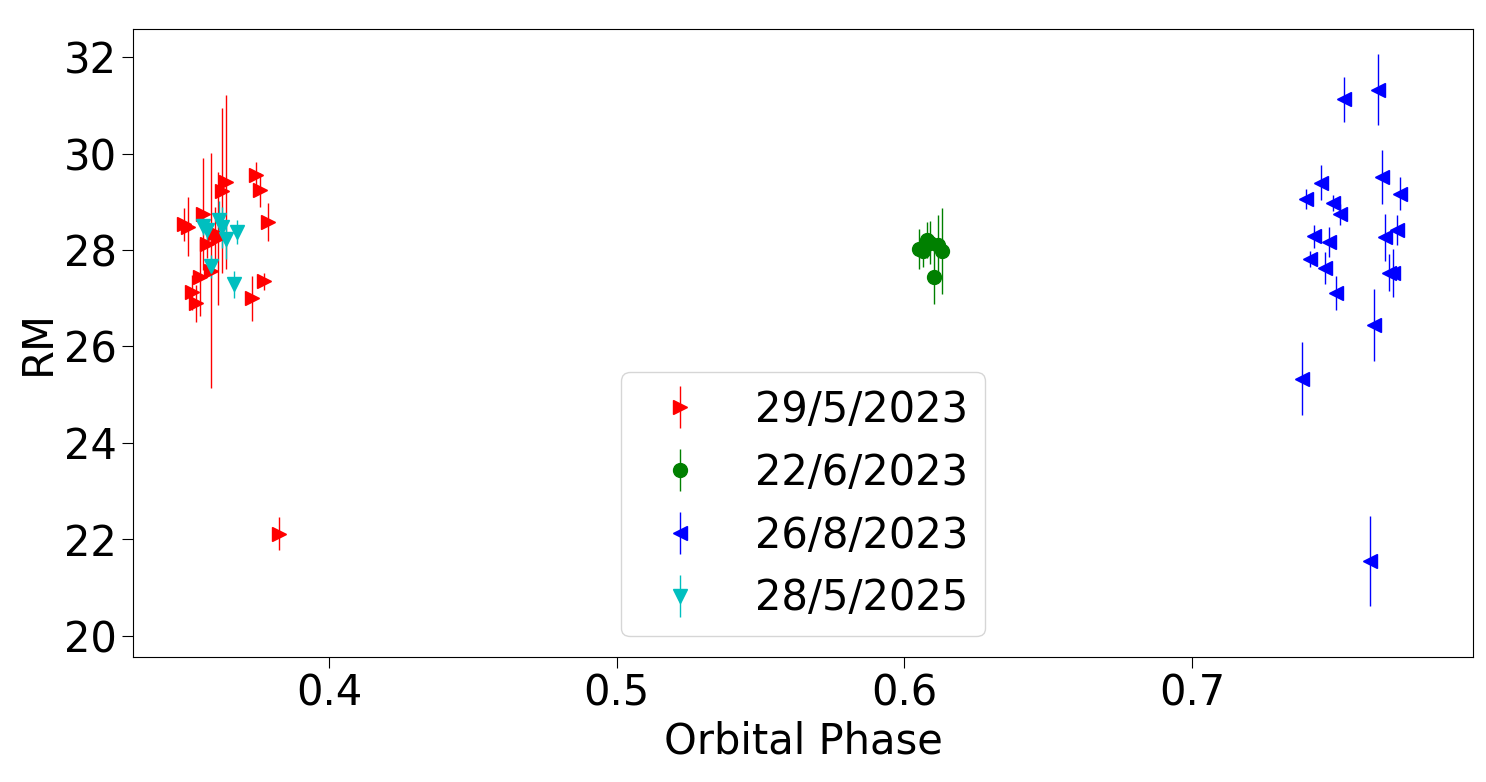}
  \caption{Temporal evolution of RM (in $rad \: m^{-2}$) across the orbital phase for all observing epochs.}
  \label{RM_poln_work}
\end{figure}

\section{Results}
\label{Results}

\subsection{RM variation with orbital phase}
\label{RM_orbital_phase}
We report RM variation (from 22 to 31 $rad \: m^{-2}$) across the orbital phase estimated for the 20 minute time chunks plotted in Figure \ref{RM_poln_work}. 
We note that accurate RM determination is crucial and the inaccuracies in RM estimation can propagate into errors in derived Stokes $Q$, $U$ parameters, which will in turn affect the computed spectra of these components. \par
No systematic trend is evident in the variation of the observed RM with orbital phase. Improved orbital phase coverage with higher cadence, would enable to us identify the possible trend in estimated RM with orbital phase. The observed variation of RM, can be attributed to fluctuations in electron density ($n_e$) and parallel component of the magnetic field to the line of sight ($B_{||}$).
\subsection{Spectro-polarimetric analysis of Stokes parameters}
In this section, we present the results from the spectro-polarimetric analysis represented in Figures \ref{All_comp_spectra_fig} to \ref{Main_comp_spectra_1_fig}. The $1^{st}$ row is a series of 2d plots which shows variation in Stokes $I$, $Q$, $U$, $V$, and $L$ with time (in x-axis; $TC_S$, $TC_M$, $TC_E$ representing the start, middle and end time chunk, each time chunk with 20 minutes duration) and frequency (in y-axis). The $2^{nd}$ row plots the time and frequency averaged profiles for each of the Stokes parameters. Time-averaged spectral information of the initial, middle and end time chunk (each time chunk with 20 minutes duration) for a given epoch are presented in the $3^{rd}$, $4^{th}$, and $5^{th}$ rows. Each point in the spectra were computed over a frequency width of 64 MHz.\par
We present spectro-polarimetric analysis, for all the profile components combined, observed on 22 June 2023, in Figure \ref{All_comp_spectra_fig}. Other epochs did not show high SNR signal for the all the components in Stokes $Q$, $U$ and $V$ parameters. We also present the similar results considering only the main profile component, marked in blue dots in $2^{nd}$ row $1^{st}$ column of Figures \ref{Main_comp_spectra_2_fig} and \ref{Main_comp_spectra_1_fig}, observed on epochs 22 June 2023 and 29 May 2023 respectively. Similar Figures for 26 August 2023 and 28 May 2025 can be found on zenodo link (\url{https://doi.org/10.5281/zenodo.16539208}).\par 
Previous studies have shown that for some pulsars, Stokes  parameters ($Q$, $U$, and $V$) changes from positive to negative values across the pulse phase bins \citep[e.g. PSR J1017$-$7156 presented in Figure A3 of ][]{Dai_et_al_2015}. For such pulsars, averaging over the pulse phase, to estimate spectra, can lead to depolarization. However, in J2144$-$5237, we do not observe sign-changing of Stokes parameters along pulse phase, which is evident in Stokes profiles shown in the $2^{nd}$ row of Figures \ref{Main_comp_spectra_2_fig}, and \ref{Main_comp_spectra_1_fig}. Therefore, averaging on pulse phase is not expected to depolarize the Stokes parameter spectra for J2144$-$5237.\par
Finally, the last column present the Poincar\'e sphere representation of the all the components (as shown by dots in the $2^{nd}$ row) for the subintegration of 20 minutes. To facilitate this interpretation of polarization variations with the pulse phase, the same colored dots, representing a particular phase bin, are overlaid on the Stokes $I$ and $L$ profiles (columns 1 and 5, respectively) and the Poincar\'e sphere. This serves as a reference for mapping specific phase bins onto the Poincar\'e sphere.

\subsubsection{Spectro-polarimetric results for the full profile}
\label{All_comp_spectra}
\begin{figure*}[htbp]
  \centering
  \vspace*{-1cm}
  \hspace*{-0.39cm}
  \resizebox{0.28\height}{0.35\width}{\rotatebox[units=-360,]{90}{\includegraphics{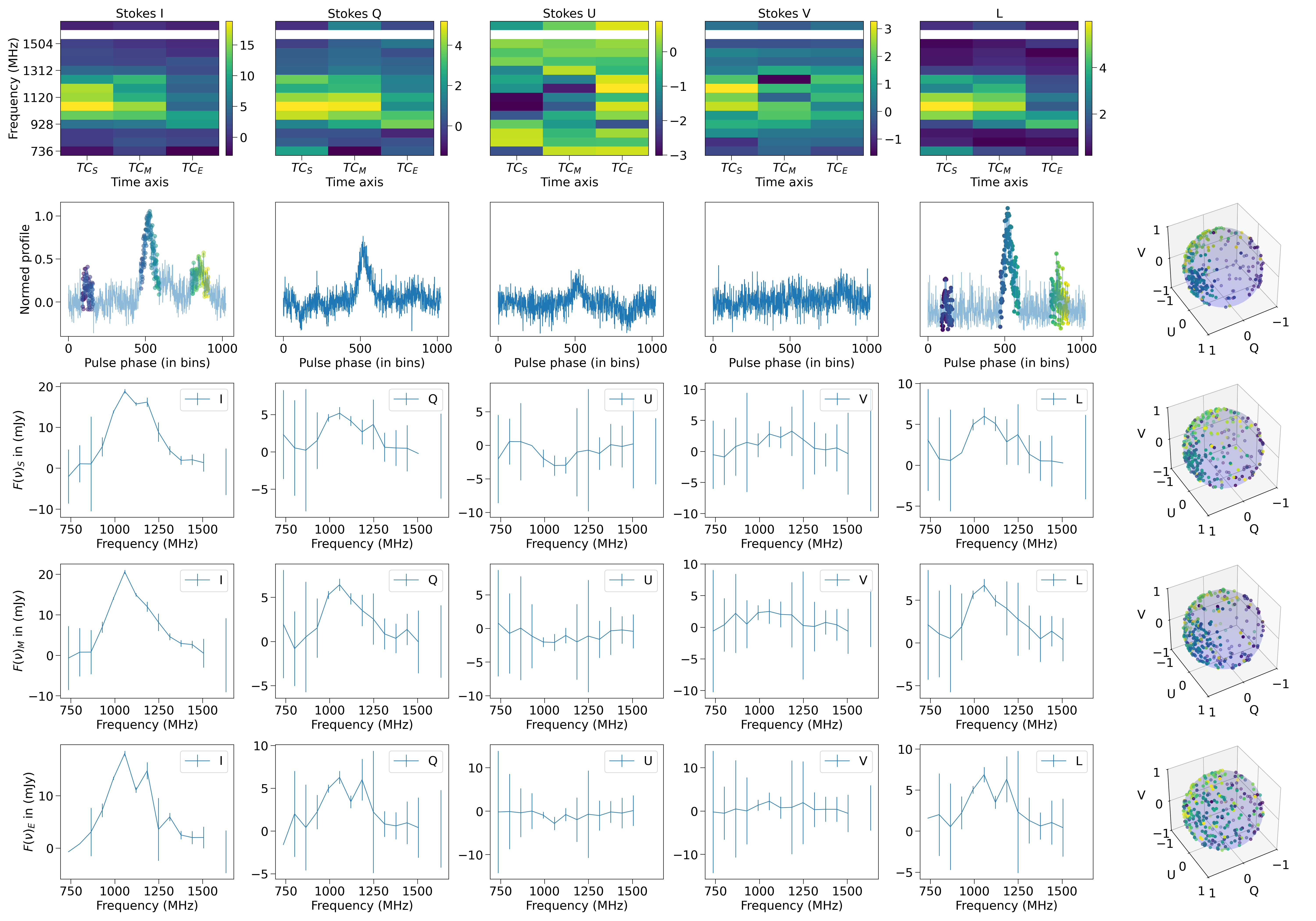}}}
  \caption{Results from Spectro-polarimetric analysis of Parkes UWL data for PSR J2144$-$5237 considering all the components (marked in coloured dots in the $2^{nd}$ row of the Figure). The columns display Stokes parameters ($I$, $Q$, $U$, $V$), linear polarization ($L$), and a Poincar\'e representation of Stokes $Q$, $U$, and $V$. The $1^{st}$ row shows spectral-temporal variations for start, middle, and end time chunks denoted as $TC_S$, $TC_M$ and $TC_E$, while the $2^{nd}$ row provides time-frequency averaged Stokes profiles and the corresponding Poincar\'e sphere representation. The coloured dots, in the $2^{nd}$ row, in the Stokes $I$ and $L$ profiles (columns 1 and 5, respectively) serve as a visual aid for mapping the phase bin points onto the Poincar\'e sphere. The last three rows display time-averaged spectra for three different time chunks ($F(\nu)_S$, $F(\nu)_M$,$F(\nu)_E$ corresponds to the flux density at the start, middle and end time chunks of the observing epoch) and the corresponding Poincar\'e sphere representations.} 
  \label{All_comp_spectra_fig}
\end{figure*}
Spectro-polarimetric results for the observing epoch 22 June 2023, of Stokes $I$, $Q$, $U$, $V$ and $L$ along with Poincar\'e sphere representation of the Stokes parameters are presented in the Figure \ref{All_comp_spectra_fig}. 
In the $2^{nd}$ row, both the main component and the adjacent component (the component situated right next to the main component which intersects with the main component spanning from around 550$-$650 phase bins) exhibit significant Stokes $I$ signal but other Stokes parameters ($Q$, $U$ and $V$) and linear polarization $L$ are not detected in the adjecent component. 

The spectra of total intensity, plotted in the $1^{st}$ column (across last three rows of Figure \ref{All_comp_spectra_fig}) display evidence of a Gigahertz turnover around 1.2 GHz. At 300 MHz, the scintillation bandwidth is approximately 45 kHz, as reported in \cite{Sharan_et_al_2024}. Applying a frequency dependence of $\nu^{-4}$ \citep{LorimerKramer} yields an estimated scintillation bandwidth of about 11 MHz at 1.2 GHz, which is smaller than the 64 MHz frequency resolution used in our spectra. 

A similar turnover trend around Gigahertz frequencies is observed in the  spectra of linear polarization ($L$) (plotted in column 5 of Figure \ref{All_comp_spectra_fig}). We noted the appearance of a spectral turnover in certain epochs, but not in others. A turnover can only be identified when both the rising (pre-turnover) and declining (post-turnover) portions of the spectrum are sampled within the $700$ MHz $–$ $4$ GHz range, or within the portion of the data that remains unflagged. Since there is a possibility of an orbital phase dependence of the observed turnover frequency (e.g. \cite{Kijak_et_al_2011}), higher cadence sampling of the orbital phase is required for further investigation. \par
Spectra with such turnovers are commonly referred to as Gigahertz peaked spectra (GPS). GPS behaviour has been reported in some pulsars, which could be linked to propagation effects, magnetospheric effects, binary interactions, supernova remnant or pulsar wind nebulae interaction with the surrounding medium \citep{Kijak_et_al_2011, Karolina_et_al_2021}. However, to our knowledge, this is the first evidence for observations of the polarization nature of GPS for any MSP. Identifying full polar GPS behavior in MSPs can provide insights into their evolutionary pathways and binary interactions \citep{Dembska_at_al_2014}.\par
The Poincar\'e sphere representation of Stokes $Q$, $U$, and $V$ (presented in the last column of the $2^{nd}$ row) shows that adjacent pulse phase bins map to nearby points in Poincar\'e sphere as well. This indicates that, profile components are clustered in the Poincar\'e sphere.

\subsubsection{Spectro-polarimetric analysis of the main component of the profile}
\label{Main_comp_spectra_2}
\begin{figure*}[htbp]
  \centering
  \vspace*{-1cm}
  \hspace*{-0.5cm}
  \resizebox{0.3\height}{0.4\width}{\rotatebox[units=-360,]{90}{\includegraphics{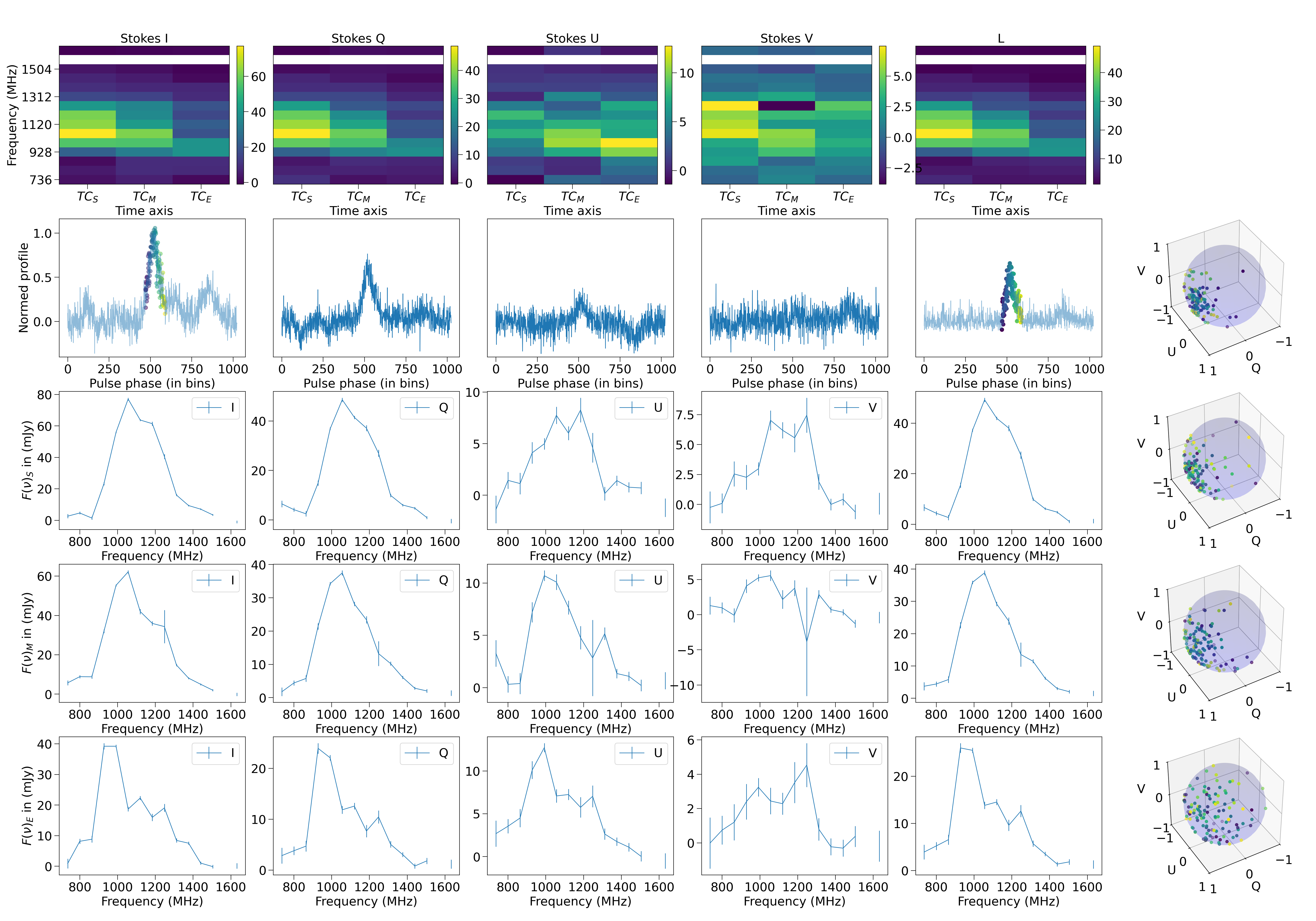}}}
  \caption{The Figure presents the Spectro-polarimetric results obtained from analysis of Parkes UWL data for PSR J2144$-$5237 observed on 22 June 2023, only for the main component. The Figure description is same as Figure \ref{All_comp_spectra_fig}.}
  \label{Main_comp_spectra_2_fig}
\end{figure*}

\begin{figure*}[htbp]
  \centering
  \vspace*{-1cm}
  \hspace*{-0.5cm}
  \resizebox{0.3\height}{0.4\width}{\rotatebox[units=-360,]{90}{\includegraphics{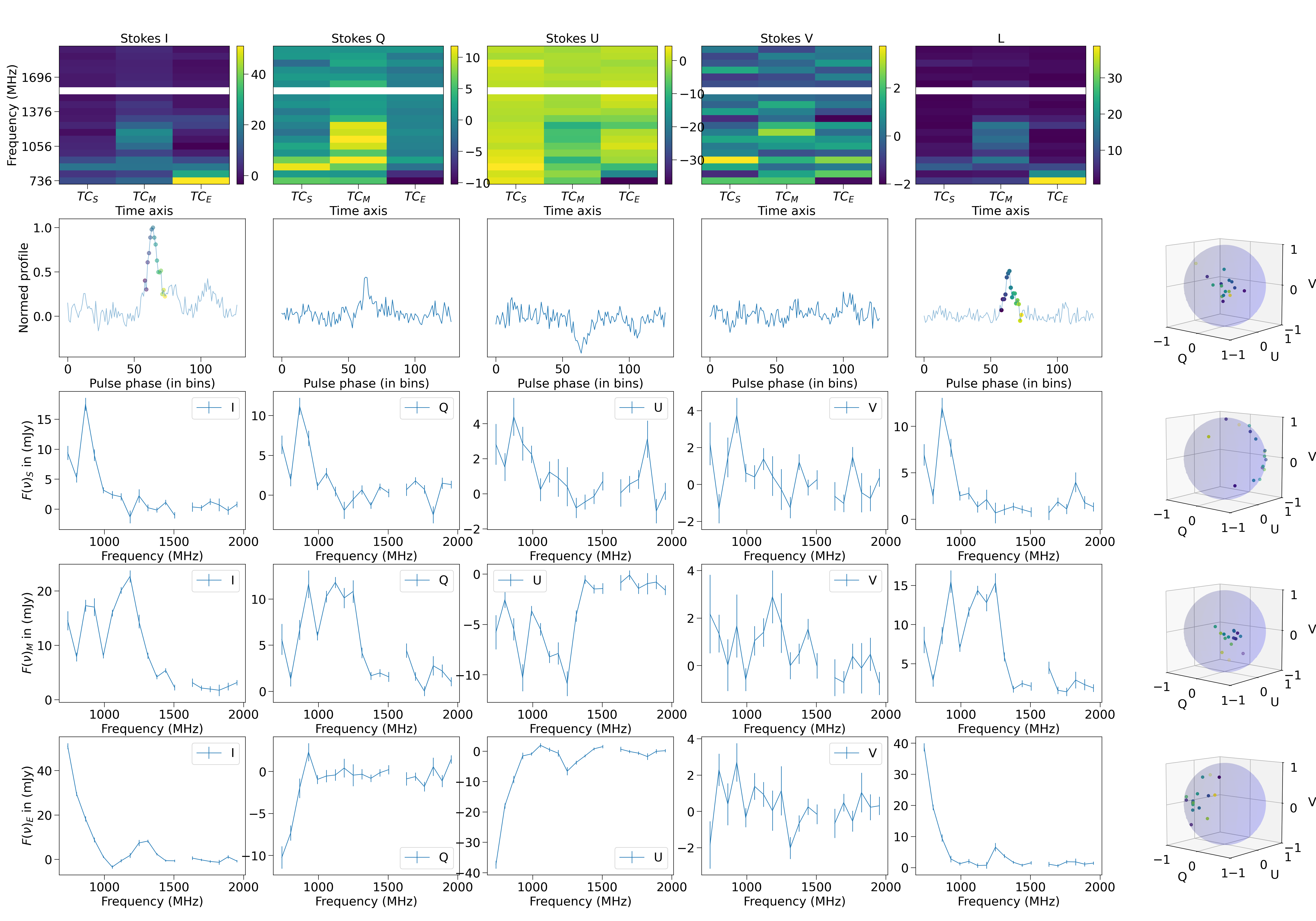}}}
  \caption{The Figure presents the Spectro-polarimetric results obtained from analysis of Parkes UWL data for PSR J2144$-$5237 observed on 29 May 2023, only for the main component. The Figure description is same as Figure \ref{All_comp_spectra_fig}. Since this epoch has low SNR for polarized signal, an average along phase bin axis by a factor of 8 was taken which results into 128 phase bins.}
  \label{Main_comp_spectra_1_fig}
\end{figure*}

Since J2144$-$5237 have a relatively large duty cycle with 5 components in the Stokes I profile, the spectro-polarimetric analysis, presented in Section \ref{All_comp_spectra}, will have contributions from all these components. To better isolate the influence of the binary medium and minimize potential contamination from magnetospheric effects associated with other components, the analysis was refined to focus exclusively on the main component of the Stokes $I$ profile. The selected phase bins representing the main component are marked by dots in the $2^{nd}$ row, $1^{st}$ column of Figure \ref{Main_comp_spectra_2_fig}, and are used for subsequent analysis.\par
The spectral and temporal variations of the Stokes parameters ($I$, $Q$, $U$, $V$) and linear polarization ($L$) are plotted in the $1^{st}$ row of Figure \ref{Main_comp_spectra_2_fig}. Visual inspection of these two-dimensional plots reveals, stokes $I$ and $Q$, as well as the linear polarization $L$, appears to decrease with time (i.e. in two dimensional plots, the $1^{st}$ column pixel values decrease from a higher value to a lower value along the time axis). These variations offer insight into the mechanisms influencing the behavior that emerges on timescales longer than 20 minutes in the polarized emission from the pulsar. Visual inspection of these two-dimensional plots also reveals a strong correlation between Stokes $I$, $Q$, and $L$, a moderate correlation between $I$ and $V$, and a weaker correlation between $I$ and $U$. \par
Consistent with the analysis including all components (Figure \ref{All_comp_spectra_fig}), the main component also exhibits a GPS and a change of turnover frequency evident for Stokes $I$, $Q$ and $L$ (as seen in last 3 rows of Figure \ref{Main_comp_spectra_2_fig}).\par
Figure \ref{Main_comp_spectra_1_fig} presents the spectro-polarimetric analysis of the main component from the observation on 29 May 2023. Towards the beginning of the observations, Stokes $U$ shows high correlation with Stokes $I$. In contrast, towards the end of observations, Stokes $U$ becomes anti-correlation with Stokes $I$ (Figure \ref{Main_comp_spectra_1_fig}), indicating a change in the polarization behavior over time.

The other interesting feature to note is the location of scattered points (corresponding to the main component) in Poincar\'e sphere, observed in 29 May 2023 epoch, drifting in a systematics manner. In the Poincar\'e sphere representation shown in the last column of Figure \ref{Main_comp_spectra_1_fig}, the average of the three subintegrations, (top pannel), appears to lie approximately close to the middle subintegration ($2^{nd}$ row from the bottom). This suggests the scatter point drifts in a systematic manner, in the sense that while performing the average, the drift between first and middle subintegration is canceled by drift between middle and the last subintegration. Poincaré sphere representations for all subintegrations from the epochs of 22 June 2023 and 29 May 2023 are provided in the Appendix.

\subsection{Correlation coefficient of Stokes parameters spectra}
\label{CC_CC_IQUV}

\begin{figure*}[htbp]
  \centering
  \includegraphics[scale=0.32] {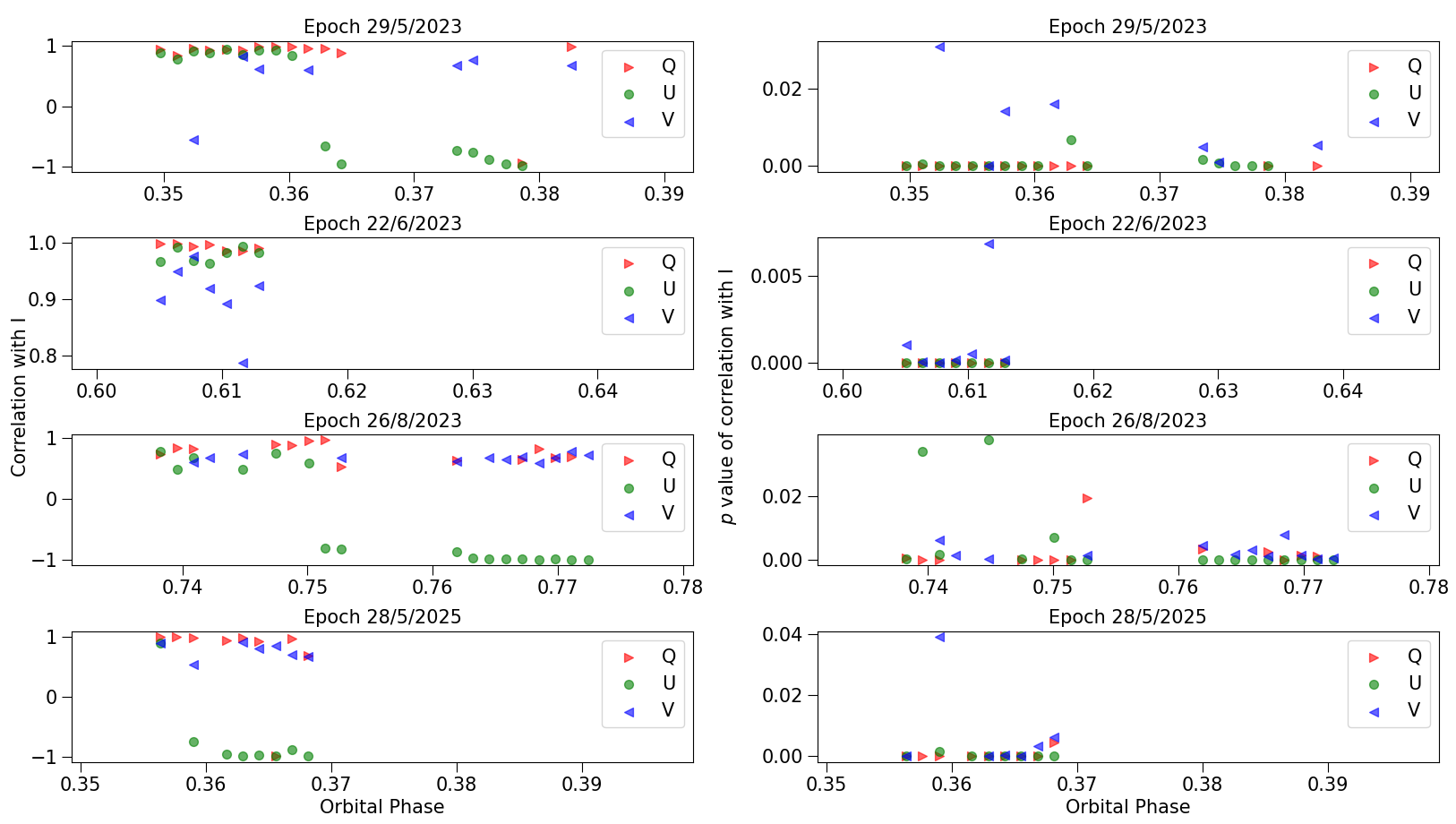}
  \caption{The figure illustrates the Pearson correlation coefficient along with corresponding $p$ values. We plot the correlation for which the $p$ values are less than 0.05}
  \label{Pearson_correlation}
\end{figure*}

We estimate Pearson correlation coefficient for the main component, which represents covariance of Stokes $I$ and other Stokes parameter ($Q$, $U$, $V$) divided by the product of their standard deviations. We attempted to use the ratios $Q/I$, $U/I$, and $V/I$, but in the presence of scintillation, these quantities show substantial fluctuations, making it difficult to infer any meaningful trends in the polarization properties. Figure \ref{Pearson_correlation} shows the estimated Pearson correlation values as a function of orbital phase, only for the subintegrations with a $p$ value less than $0.05$.\par
A strong correlation coefficient, between Stokes $I$ and the rest of the Stokes parameters for the main component, can be noted in orbital phase from 0.35 to 0.36, from 0.605 to 0.613, and 0.74 to 0.75 (covered on 29 May 2023 , 28 May 2025, 22 June 2023, and 26 August 2023 respectively) as seen in Figure \ref{Pearson_correlation} (mentioned in Section \ref{Main_comp_spectra_2}). 
In epoch 26 August 2023, positive correlation in spectra of Stokes $I$ and $U$ is observed till orbital phase up to 0.75 and then converts to high anti-correlation from orbital phase 0.75 to 0.773, last row of Figure \ref{Pearson_correlation}. Similarly, for epochs 29 May 2023 and 28 May 2025, positive correlation (between spectra of Stokes $I$ and $U$) till orbital phase 0.36 is observed and then it drastically changes to anti-correlation ($1^{st}$ row of Figure \ref{Pearson_correlation}). For correlation coefficient between Stokes $I$ and $Q$ (or $V$), in general positive correlation is observed across all the orbital phase. Additional observations covering the full orbital phase of this binary system will be required to decipher the possible orbital phase dependent patterns in these correlation coefficients.\par
The $p$ value is crucial in understanding if the observed correlation between Stokes parameters are due to some physical phenomena. Higher the $p$ value higher is the chance that the correlation coefficient calculated is one of the realization of random sampling and is not by some physical phenomena. If the $p$ value is lower then the probability of getting the observed correlation value by chance is lower. In words, low p-value is due the correlation caused by some physical phenomena.\par
As one can note from the correlation coefficient plots (Figure \ref{Pearson_correlation}), a systematic but sudden change, from positive to negative value at around orbital phase 0.36 (at two epochs) and 0.75, is observed. Such sudden change of polarization could be attributed interaction between the radio signal generated from the pulsar with magnetized plasma in the intra-binary material.  

\section{Summary}
\label{summary}
Investigation of spectra and polarization are two fundamental blocks for understanding the properties of the MSPs. In this work, we combine these two components and present a method for the first time-varying investigation of the spectro-polarimetric properties for any pulsar so far. We have successfully applied this method on the MSP J2144$-$5237, and the main new results from this work are listed below,
\begin{itemize}
    \item The range of RM values, across the orbital phase, from 22 to 31 is reported in this study, without any significant dependence with orbital phase. This variation of RM could partially be due to the intra-binary material. Additional observation and hence larger sampling of RM measurements across binary orbit can provide more information of distribution of material and magnetic field orientation. Orbital phase dependent RM variation is reported only for a few of other compact orbit MSPs \citep{Kumari_et_al_2024a, Polzin_et_al_2019, Crowter_et_al_2020, Wang_et_al_2023b}.
    \item We have observed temporal evolution of the spectra for MSP J2144$-$5237, with a time resolution of 20 minutes. The temporal changes in flux density modulation in the spectra of Stokes $I$, $Q$, and circular polarization $V$ is observed to be correlated on all orbital phase. Whereas we observe Stokes $U$ highly correlated with Stokes $I$ at orbital phase of some parts of the orbital phase but anti correlated at other orbital phase. For example, the correlation for epochs 29 May 2023 and 28 May 2025 indicates a transition occurring at nearly identical orbits phase.
    \item For the first time we report polarization properties for GPS in any MSP. We also report the drift in the GPS turnover frequency with orbital phase for this MSP. Previous studies, on other binary pulsars, reported the change in spectra of Stokes $I$ \citep{Dembska_at_al_2014, Karolina_et_al_2021, Kijak_et_al_2011}. For example in \cite{Kijak_et_al_2011}, authors have reported spectral evolution of B1259$-$63 with orbital phase. Such studies show that GPS pulsars are often situated in complex environments, such as pulsar wind nebulae or supernova remnants, where thermal free-free absorption or synchrotron self-absorption mechanisms are believed to cause the observed spectral turnovers \citep{Rajwade_et_al_2016}.
    \item A Poincar\'e sphere representation of the Stokes $Q$, $U$ and $V$ is presented, showing nearby phase bins are clustered together suggesting that the polarization state changes relatively slowly with the pulse phase. The trajectory of points on the Poincar\'e sphere reveals how polarization properties changes across the pulse phase. For two of the epochs (29 May 2023 and 26 August 2023), we observe the systematic change of the cluster of points, of main component phases bins, on Poincar\'e sphere. 
\end{itemize}

In future, we aim to probe possibility of depolarization near the superior conjunction with larger samplings of RM across orbital phase, which was not probed in the current study. There are observational evidence of depolarization near superior conjunction for spider MSPs such as PSR J1748$-$2446A \citep{Li_et_al_2023}, J2051$-$0827 \citep{Polzin_et_al_2019, Wang_et_al_2023b}, and J2256$-$1024 \citep{Crowter_et_al_2020}.

We also aim for investigation of spectro-polarimetric properties of J2144$-$5237 at highly sampled orbital phase which will help us disentangling the environmental and intrinsic effects responsible for the GHz peaked spectrum. Additionally, we would also conduct investigation of time varying nature of spectro-polarimetric properties for a bigger sample of MSPs and normal pulsars using wider frequency range (i.e. perform simultaneous observations, wherever possible, using low-frequency (LOFAR, uGMRT), mid-frequency (uGMRT, MeerKat) and high-frequency (Meerkat, Effelsberg) radio telescopes). \par
Thus in summary, we present new results from a hitherto unexplored regime of temporal changes in spectro-polarimetric properties of MSP J2144$-$5237 and will be applied to a larger sample of pulsars.

\section{Acknowledgements}
We thank the Department of Atomic Energy, Government of India, under project No.12-R\&D-TFR-5.02-0700.The data presented in this paper were obtained by the Parkes radio telescope. Murriyang, CSIRO’s Parkes radio telescope, is part of the Australia Telescope National Facility (https://ror.org/05qajvd42) which is funded by the Australian Government for operation as a National Facility managed by CSIRO. We acknowledge the Wiradjuri people as the Traditional Owners of the Observatory site. Lastly, we acknowledge the discussions with Apurba Bera regarding Poincar\'e sphere representation of Stokes parameters.

\bibliography{sample631.bib}{}
\bibliographystyle{aasjournal}

\software{astropy \citep{astropy:2013, astropy:2018, astropy:2022},  
          PSRPYPE \citep{PSRPYPE}}

\appendix
\label{appendix}
\renewcommand\thesection{A}
\renewcommand\thefigure{\thesection.\arabic{figure}}    
\setcounter{figure}{0}    

\begin{figure*}[htbp]
  \centering
  \includegraphics[scale=0.8]{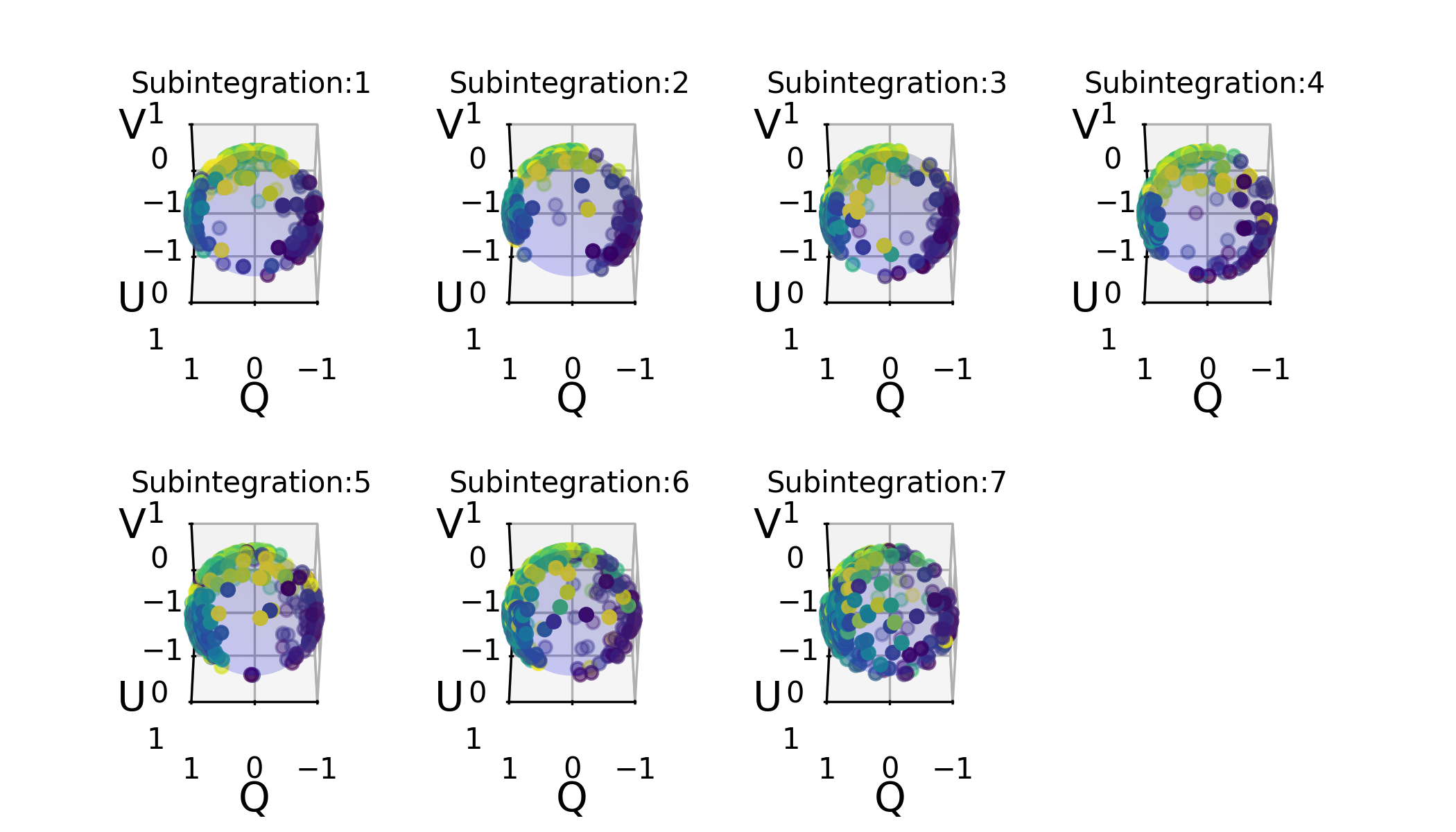}
  \caption{The figure presents the Poincaré sphere variation with subintegrations obtained from analysis of PSR J2144$-$5237 observed on 22 June 2023, all the components. The colour scheme of the points in the figure is same as of Figure \ref{All_comp_spectra_fig}.}
  \label{Poincare_sphere_E2}
\end{figure*}

\begin{figure*}[htbp]
  \centering
  \includegraphics[scale=0.8]{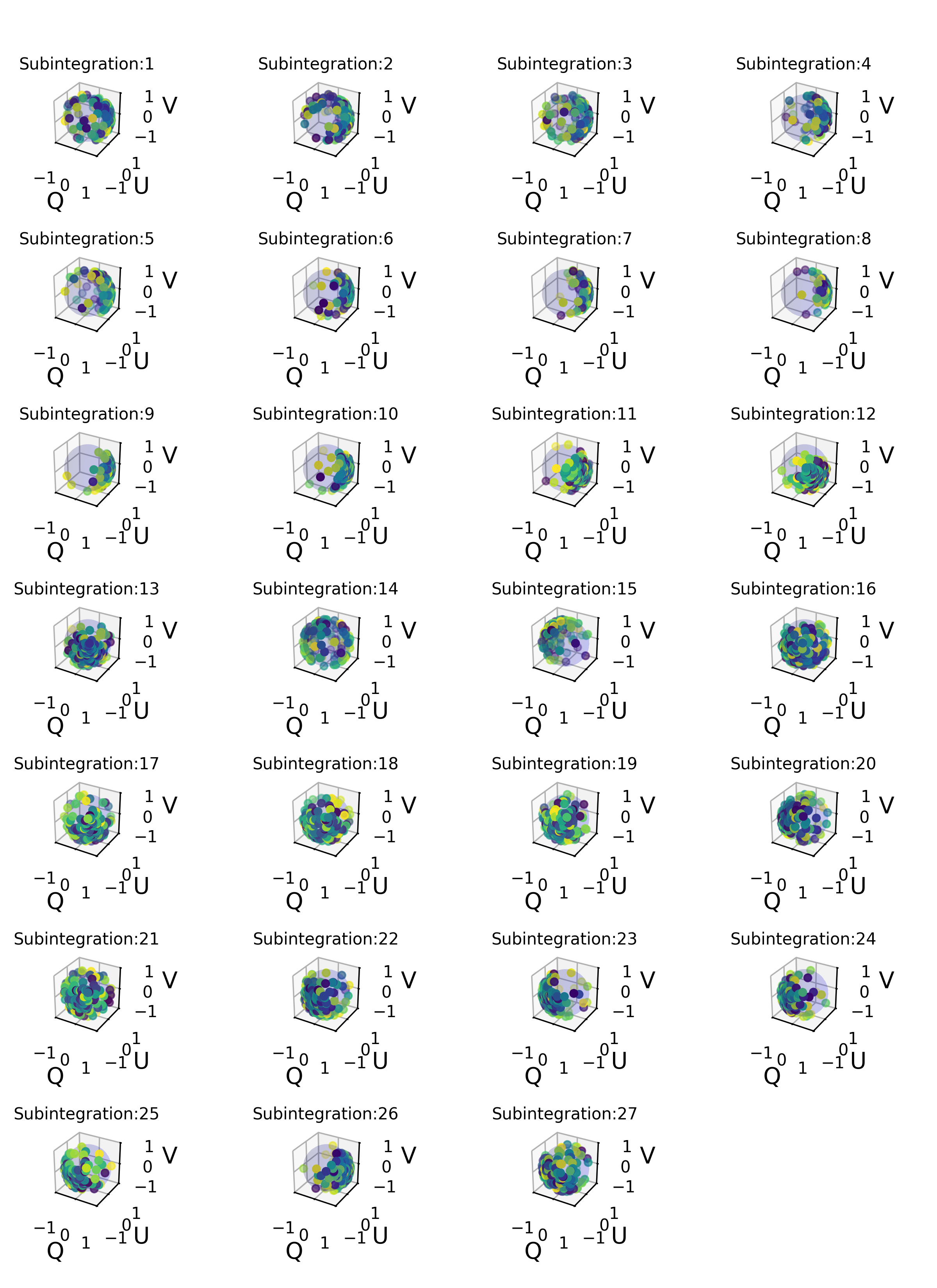}
  \caption{The figure presents the Poincaré sphere variation with subintegrations for observation on 29 May 2023, only for the main component. The colour scheme followed is same as in Figure \ref{Main_comp_spectra_2_fig}.}
  \label{Poincare_sphere_E1}
\end{figure*}

We present the spin phase variation of Poincaré sphere representation of the Stokes $Q$, $U$, and $V$ parameters with different subintegrations (which represents different orbital phases). \textit{As a powerful tool for visualizing component-dependent differences, the Poincaré sphere variation across subintegrations helps to quantify polarization changes with orbital phase.}
Figure \ref{Poincare_sphere_E2} shows variation Poincaré sphere representation for all the $3$ components, across different subintegrations (representing orbital phases), observed on 22 June 2023. Different components are clustered at different part of the Poincaré sphere. The variation between different panels, each panel being the 20 minutes time averaged plot, represents the systematic change of component dependent Poincaré sphere representation of Stokes parameters with orbital phases. Figure \ref{Poincare_sphere_E1}, shows similar variation of Poincaré sphere representation, with orbital phase only for the main component, observed on 29 May 2023. The change of Stokes $U$ from $+$ve to $-$ve value, across different subintegrations (described by different panels), is evident from the Figure \ref{Poincare_sphere_E1}. On the other hand, 22 June 2023 epoch (presented in Figure \ref{Poincare_sphere_E2}) does not show systematic changes of Stokes parameters values in the Poincaré sphere representation across different subintegrations (or orbital phase), which is also evident from the Pearson correlation coefficient plots shown in Figure \ref{Pearson_correlation}. Similar Poincaré sphere representations for other epochs (i.e. 26 August 2023 and 28 May 2025) can be found on zenodo link (\url{https://doi.org/10.5281/zenodo.16539208}).

\end{document}